\documentclass[reprint, twocolumn, showkeys]{revtex4}

\usepackage{graphicx}% Include figure files
\usepackage{dcolumn}% Align table columns on decimal point
\usepackage{bm}% bold math
\usepackage{color}% bold math
\usepackage{amsmath,amsthm}
\usepackage{amssymb,bm}
\usepackage{mathrsfs}
\usepackage{comment}
\usepackage{ulem}

\graphicspath{%
    {converted_graphics/}% inserted by PCTeX
    {/}% inserted by PCTeX
}
\begin{document}

\title{Semiclassical truncated-Wigner-approximation theory of molecular-vibration-polariton dynamics in optical cavities}
\author{Nguyen Thanh Phuc}
\email{nthanhphuc@moleng.kyoto-u.ac.jp}
\affiliation{Department of Molecular Engineering, Graduate School of Engineering, Kyoto University, Kyoto 615-8510, Japan}

%%%%%%%%%%%%%%%%%%%%%%%%%%
\begin{abstract}
It has been experimentally demonstrated that molecular-vibration polaritons formed by strong coupling of a molecular vibration to an infrared cavity mode can significantly modify the physical properties and chemical reactivity of various molecular systems. 
However, a complete theoretical understanding of the underlying mechanisms of the modifications remains elusive due to the complexity of the hybrid system, especially the collective nature of polaritonic states in systems containing many molecules. We develop here the semiclassical theory of molecular-vibration-polariton dynamics based on the truncated Wigner approximation (TWA) that is tractable in large molecular systems and simultaneously captures the quantum character of photons in the optical cavity. 
The theory is then applied to investigate the nuclear quantum dynamics of a system of identical diatomic molecules having the ground-state Morse potential and strongly coupled to an infrared cavity mode in the ultrastrong coupling regime. 
The validity of TWA is examined by comparing with the fully quantum dynamics of a single-molecule system for two different initial states in the dipole and Coulomb gauges. 
For the initial tensor-product ground state in the dipole gauge, which corresponds to a light-matter entangled state in the Coulomb gauge, the collective and resonance effects of molecular-vibration-polariton formation on the nuclear dynamics are observed in a system of many molecules.  
\end{abstract}


\maketitle

%%%%%%%%%%%%%%%%%%%%%
%\section{TOC Graphic}
%\begin{figure}[h] % float placement: (h)ere, page (t)op, page (b)ottom, other (p)age
  %\centering
  % file name: D:/Draft- Chiral Cavity Induced Spin Selectivity (Sep 2022)/TOC.eps
  %\includegraphics[width=3.25in, keepaspectratio]{TOC}
    %\label{fig:TOC}
%\end{figure}

\textit{Introduction--}
Strongly coupling a molecular-vibration mode to an infrared cavity mode results in the formation of molecular-vibration polaritons--hybrid states with both light and matter characters. 
Numerous experiments have demonstrated that strong light-matter coupling can significantly modify the physical properties and chemical reactivity of molecular systems~\cite{Nagarajan21, Simpkins23, Hirai23}. 
Examples include the acceleration or hindrance of the chemical reactivity of molecules~\cite{Thomas16, Vergauwe19, Lather19, Hirai20, Pang20, Sau21, Lather21}, chemoselectivity~\cite{Thomas19}, supramolecular assembly and crystallization~\cite{Joseph21, Hirai21}, vibrational energy transfer~\cite{Xiang20}, ferromagnetism, and superconductivity~\cite{Thomas21, Thomas19b}.
Despite great efforts~\cite{Galego19, Angulo19, Phuc20, Li20, Li21, Schafer22, Du22, Riso22, Sun22, Wang22, Wang22b, Lindoy23, Fiechter23}, the mechanisms underlying changes to molecular properties and reactivity from strong vibrational coupling remain unclear~\cite{Wang21, Fregoni22, Angulo23, Mandal23, Ruggenthaler23}. 
One of the difficulties is due to the inherently collective nature of molecular polaritons in systems containing many molecules while performing a fully quantum mechanical simulation of molecular dynamics is prohibitively expensive for such a complex hybrid system.

On the other hand, it is possible for strong light-matter coupling to occur even in the absence of light due to the quantum nature of the electromagnetic field.
The optical mode has zero-point energy, which results in vacuum fluctuations.
In this work, we develop a semiclassical theory of molecular-vibration-polariton dynamics that is applicable to large molecular systems while simultaneously capturing the quantum character of photons in the optical cavity.
It is based on the truncated Wigner approximation (TWA), in which the equation of motion for the Wigner function in the phase-space representation is truncated at the leading order in the expansion with respect to the parameter characterizing the quantum fluctuation~\cite{Moyal49, Hillery84, Polkovnikov10}.
The leading-order equation of motion is unaffected by quantum fluctuations, which are only reflected in the Wigner distribution of the initial state.
In other words, in the TWA, the Heisenberg uncertainty principle for the quantum fluctuations is satisfied, while the equation of motion of physical observables in the phase space takes the classical form.
The TWA theory has been developed using quantum coherent states for applications in quantum optics and bosonic cold atoms~\cite{Walls-book, Blakie08} and using the coordinate-momentum representation in the context of quantum dynamics~\cite{Hillery84, Zurek03}.
In this work, we combine the coherent-state and coordinate-momentum representations to develop the TWA theory for strongly coupled molecule-cavity systems.

The developed theory is then applied to investigate the nuclear quantum dynamics of a system of identical diatomic molecules. 
These molecules have a ground-state Morse potential~\cite{Mondal23} and they are strongly coupled to an infrared cavity mode in the ultrastrong coupling regime, where the molecule-cavity coupling strength is comparable to both the cavity and molecular vibrational frequencies~\cite{Kockum19}. 
As a result of the ultrastrong coupling, both the molecular nuclei and cavity field evolve with time even if they are initially prepared in their ground states.
To examine the validity of the TWA, a comparison with the fully quantum dynamics obtained by solving the quantum master equation is made for a single-molecule system.
For the initial state that is a tensor product of the molecular vibrational ground state and the photonic vacuum state in the dipole gauge and corresponds to a light-matter entangled state in the Coulomb gauge, a good agreement between semiclassical and quantum mechanical results is obtained.
For a system of many molecules, the nuclear dynamic amplitude does not decrease with the increasing number of molecules, provided that the collective coupling strength is kept constant. 
This collective effect can be attributed to the nonzero molecular dipole moment at the equilibrium point that exists in the light-matter interaction Hamiltonian in the dipole gauge, or equivalently, in the initial light-matter entangled state in the Coulomb gauge.
The collective and resonance effects of molecular-vibration-polariton formation on the nuclear dynamics are also observed in a system of molecules with random orientations.
By contrast, for the initial tensor-product ground state in the Coulomb gauge corresponding to a light-matter entangled state in the dipole gauge, a noticeable difference between the semiclassical TWA and quantum mechanical results is observed as a consequence of the strong light-matter entanglement.

\begin{figure}[tbp] % float placement: (h)ere, page (t)op, page (b)ottom, other (p)age
  \centering
  % file name: E:/Draft-Semiclassical TWA theory of molecular vibration polariton dynamics (Sep 2023)/system.eps
  \includegraphics[width=3.4in, keepaspectratio]{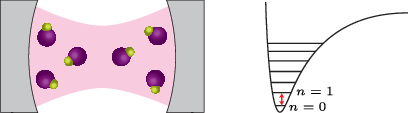}
  \caption{Schematic illustration of a system of identical diatomic molecules whose nuclear motions are strongly coupled to an infrared cavity mode (light magenta) of an optical resonator. The molecule's electronic-ground-state potential is modeled by the Morse potential, and the cavity frequency is in close resonance with the energy difference between the vibrational ground state ($n=0$) and the first excited state ($n=1$).}
  \label{fig: system}
\end{figure}

\textit{TWA theory of molecular-vibration-polariton dynamics--}
As illustrated in Fig.~\ref{fig: system}, we consider a system of $N$ molecules strongly coupled to an optical cavity mode whose frequency $\omega_\text{c}$ is in the infrared region so that the electronic excitations of the molecules can be ignored. 
The dipole-gauge total Hamiltonian of the system $\hat{H}=\hat{H}_\text{m}+\hat{H}_\text{c}$ is obtained by performing the Power-Zienau-Woolley (PZW) gauge transformation $\hat{U}=e^{-i\hat{\boldsymbol{\mu}}\cdot\hat{\mathbf{A}}}$ on the minimal coupling QED Hamiltonian in the Coulomb gauge~\cite{Tannoudji-book, Power59, Woolley74}.
Here, $\hat{\boldsymbol{\mu}}$ is the total dipole moment operator of the molecular system and $\hat{\mathbf{A}}$ is the vector potential operator of the cavity mode. 
The nuclear dynamics on the electronic-ground-state potential energy surface of the molecules are represented by the molecular Hamiltonian
\begin{align}
\hat{H}_\text{m}=\sum_{n=1}^N
\left[\sum_{j\in n} \frac{\hat{\mathbf{P}}_{nj}^2}{2M_{nj}}+\hat{V}_\text{g}^{(n)}\right]
+\sum_{n=1}^N\sum_{l>n}\hat{V}_\text{int}^{(nl)},
\end{align}
where $\hat{\mathbf{P}}_{nj}$ denotes the momentum operator of the $j$th nuclear degree of freedom in the $n$th molecule with the effective mass $M_{nj}$, $\hat{V}_\text{g}^{(n)}$ represents the electronic-ground-state potential energy of the $n$th molecule, and $\hat{V}_\text{int}^{(nl)}$ is the interaction between the $n$th and $l$th molecules.
The cavity photon energy and the molecule-cavity coupling are included in the Hamiltonian
\begin{align}
\hat{H}_\text{c}=&\hbar\omega_\text{c}\left(\hat{a}^\dagger\hat{a}+\frac{1}{2}\right)
+i\omega_\text{c} \left(\sum_{n=1}^N \hat{\boldsymbol{\mu}}_n\cdot \mathbf{A}_0\right) \left(\hat{a}^\dagger-\hat{a}\right) \nonumber\\
&+\frac{\omega_\text{c}}{\hbar}\left(\sum_{n=1}^N \hat{\boldsymbol{\mu}}_n\cdot \mathbf{A}_0\right)^2,
\label{eq: molecule-cavity coupling}
\end{align}
where $\hat{a}$ denotes the annihilation operator of a cavity photon, $\mathbf{A}_0=\sqrt{\frac{\hbar}{2\omega_\text{c}\epsilon\mathcal{V}}}\hat{\mathbf{e}}$ is the vector potential amplitude of the cavity field (with $\epsilon$, $\mathcal{V}$, and $\hat{\mathbf{e}}$ being the permittivity inside the cavity, effective cavity quantization volume, and unit vector of the cavity field polarization direction, respectively), and $\hat{\boldsymbol{\mu}}_n$ represents the electric dipole moment operator of the $n$th molecule in its electronic ground state. 
The dipole self-energy, which is the last term in Eq.~\eqref{eq: molecule-cavity coupling}, cannot be neglected in the ultrastrong coupling regime as it is comparable in magnitude with the other terms in the Hamiltonian~\cite{Kockum19, Rokaj18}. 

The cavity loss is caused by the coupling between the optical mode inside the cavity and the electromagnetic field environment outside the cavity. 
We assume that the effect of energy dissipation on the system's dynamics can be described phenomenologically under the Markov approximation by the Lindblad quantum master equation for the density operator $\hat{\rho}$~\cite{Lindblad76, Walls-book}:
\begin{align}
\frac{\text{d}\hat{\rho}}{\text{d}t}=-\frac{i}{\hbar}\left[\hat{H},\hat{\rho}\right]+\mathcal{L}_\text{c}\hat{\rho}.
\label{eq: quantum master equation}
\end{align}
Here, $\mathcal{L}_\text{c}$ is the Liouville superoperator, whose action on the density operator is given by
\begin{align}
\mathcal{L}_\text{c}\hat{\rho}=&\kappa \left(\bar{n}+1\right) \left(2\hat{a}\hat{\rho}\hat{a}^\dagger -\hat{a}^\dagger\hat{a}\hat{\rho}-\hat{\rho}\hat{a}^\dagger\hat{a}\right)\nonumber\\
&+\kappa \bar{n}\left(2\hat{a}^\dagger\hat{\rho}\hat{a}-\hat{a}\hat{a}^\dagger\hat{\rho}-\hat{\rho}\hat{a}\hat{a}^\dagger\right),
\label{eq: Liouville superoperator}
\end{align}
where $\kappa$ denotes the cavity loss rate and $\bar{n}=1/\left(e^{\hbar\omega_\text{c}/(k_\mathrm{B}T)}-1\right)$ is the average number of photons with frequency $\omega_\text{c}$ in the environment at temperature $T$.
Physically, the first line of Eq.~\eqref{eq: Liouville superoperator} describes the spontaneous and stimulated energy emissions while the second line corresponds to the energy absorption.  

To derive the semiclassical TWA theory for molecular-vibration-polariton dynamics, the Wigner quasi-probability distribution function is defined in the joint phase space of molecular nuclear motion and cavity field.
The Wigner function for the density operator $\hat{\rho}$ and the Weyl symbol of an arbitrary operator $\hat{\Omega}$ are defined by~\cite{Garrison-textbook}
\begin{align}
W(\mathbf{R},\mathbf{P},a)=&\;\hbar^\mathcal{N}\int\frac{\text{d}^\mathcal{N}\mathbf{u}}{(2\pi)^\mathcal{N}}\int\frac{\text{d}^\mathcal{N}\mathbf{v}}{(2\pi)^\mathcal{N}}\int\frac{\text{d}^2\eta}{\pi^2}
e^{i\left(\mathbf{u}\cdot\mathbf{P}+\mathbf{v}\cdot\mathbf{R}\right)}\nonumber\\
&\times e^{\eta^*a-\eta a^*}
\text{Tr}\left\{\hat{\rho}e^{-i\left(\mathbf{u}\cdot\hat{\mathbf{p}}+\mathbf{v}\cdot\hat{\mathbf{r}}\right)}
e^{\eta\hat{a}^\dagger-\eta^*\hat{a}}\right\}
\end{align}
and
\begin{align}
\Omega_\text{W}(\mathbf{R},\mathbf{P},a)=&\;\hbar^\mathcal{N}\int\frac{\text{d}^\mathcal{N}\mathbf{u}}{(2\pi)^\mathcal{N}}\int\frac{\text{d}^\mathcal{N}\mathbf{v}}{(2\pi)^\mathcal{N}}\int\frac{\text{d}^2\eta}{\pi^2}
e^{i\left(\mathbf{u}\cdot\mathbf{P}+\mathbf{v}\cdot\mathbf{R}\right)}\nonumber\\
&\times e^{\eta^*a-\eta a^*}
\text{Tr}\left\{\hat{\Omega}e^{-i\left(\mathbf{u}\cdot\hat{\mathbf{p}}+\mathbf{v}\cdot\hat{\mathbf{r}}\right)}
e^{\eta\hat{a}^\dagger-\eta^*\hat{a}}\right\},
\end{align}
respectively, where $\mathbf{R}=\left\{\mathbf{R}_{nj}\right\}$, $\mathbf{P}=\left\{\mathbf{P}_{nj}\right\}$, $\mathbf{u}=\left\{\mathbf{u}_{nj}\right\}$, $\mathbf{v}=\left\{\mathbf{v}_{nj}\right\}$, $\hat{\mathbf{r}}=\left\{\hat{\mathbf{r}}_{nj}\right\}$, and $\hat{\mathbf{p}}=\left\{\hat{\mathbf{p}}_{nj}\right\}$ denote the sets of coordinate and momentum variables and operators for all the nuclear degrees of freedom of the molecules inside the cavity, and $\mathcal{N}$ is the total number of degrees of freedom in the molecular system. 
The Weyl symbol of the product $\hat{\Omega}=\hat{\Omega}_1\hat{\Omega}_2$ of two operators can be obtained by using the Moyal product~\cite{Polkovnikov10}:
\begin{align}
\Omega_\text{W}=\Omega_{1\text{W}}\exp\left(\frac{\Lambda_\text{c}-i\hbar\Lambda}{2}\right)\Omega_{2\text{W}},
\end{align}
where 
\begin{align}
\Lambda=&\frac{\overleftarrow{\partial}}{\partial \mathbf{P}}\cdot \frac{\overrightarrow{\partial}}{\partial \mathbf{R}}
-\frac{\overleftarrow{\partial}}{\partial \mathbf{R}}\cdot \frac{\overrightarrow{\partial}}{\partial \mathbf{P}},\\
\Lambda_\text{c}=&\frac{\overleftarrow{\partial}}{\partial a}\frac{\overrightarrow{\partial}}{\partial a^*}
-\frac{\overleftarrow{\partial}}{\partial a^*}\frac{\overrightarrow{\partial}}{\partial a}.
\end{align}
Here, the left (right) arrow implies that the derivative acts on the operator on the left (right).
Similarly, the Weyl symbol of the commutator $\hat{\Omega}=\left[\hat{\Omega}_1,\hat{\Omega}_2\right]$ of two operators can be obtained by using the Moyal bracket:
\begin{align}
\Omega_\text{W}=2\Omega_{1\text{W}}\sinh\left(\frac{\Lambda_\text{c}-i\hbar\Lambda}{2}\right)\Omega_{2\text{W}}.
\end{align}
The quantum master equation~\eqref{eq: quantum master equation} then translates into a partial differential equation for the Wigner function. 
By dropping all terms containing higher-than-second-order derivatives, the obtained equation for the Wigner function is found to be
\begin{align}
\frac{\partial W}{\partial t}=&-\left(\frac{\partial H_\text{W}}{\partial \mathbf{P}}\cdot\frac{\partial W}{\partial \mathbf{R}}-\frac{\partial H_\text{W}}{\partial \mathbf{R}}\cdot\frac{\partial W}{\partial \mathbf{P}}\right)\nonumber\\
&-\frac{i}{\hbar}\left(\frac{\partial H_\text{W}}{\partial a}\frac{\partial W}{\partial a^*}-\frac{\partial H_\text{W}}{\partial a^*}\frac{\partial W}{\partial a}\right) \nonumber\\
&+\kappa \Bigg[\frac{\partial(a^*W)}{\partial a^*}
+\frac{\partial(a W)}{\partial a}+(2\bar{n}+1)\frac{\partial^2 W}{\partial a\partial a^*}\Bigg],
\end{align}
where $H_\text{W}$ is the Weyl symbol of the Hamiltonian $\hat{H}$.
By projecting onto the molecule's electronic ground-state manifold, we have
\begin{align}
\frac{\partial W}{\partial t}=&\sum_{n=1}^N\sum_{j\in n} 
\Bigg[-\frac{\mathbf{P}_{nj}}{M_{nj}}\cdot \frac{\partial W}{\partial \mathbf{R}_{nj}} + \Bigg(\frac{\partial V_\text{g}^{(n)}}{\partial \mathbf{R}_{nj}}+\sum_{l\not=n}\frac{\partial V_\text{int}^{(nl)}}{\partial \mathbf{R}_{nj}}\Bigg) \nonumber\\
&\cdot \frac{\partial W}{\partial \mathbf{P}_{nj}}\Bigg]
-i\omega_\text{c}\Bigg[\frac{\partial(a^* W)}{\partial a^*}-\frac{\partial(a W)}{\partial a}\Bigg]
-\frac{\omega_\text{c}}{\hbar}\sum_{n=1}^N \nonumber\\
&\Bigg[(\boldsymbol{\mu}_n \cdot \mathbf{A}_0) \Bigg(\frac{\partial W}{\partial a^*}+\frac{\partial W}{\partial a}\Bigg)
+i\hbar(a-a^*)\sum_{j\in n}  \nonumber\\
&\sum_{\lambda,\nu=x,y,z}\frac{\partial \mu_n^\lambda}{\partial R_{nj}^\nu}A_0^\lambda \frac{\partial W}{\partial P_{nj}^\nu} \Bigg]
+\frac{2\omega_\text{c}}{\hbar}\Bigg(\sum_{n=1}^N \boldsymbol{\mu}_n\cdot\mathbf{A}_0\Bigg) \nonumber\\
&\times \sum_{n=1}^N \sum_{j\in n}\sum_{\lambda,\nu=x,y,z} \frac{\partial \mu_n^\lambda}{\partial R_{nj}^\nu}A_0^\lambda \frac{\partial W}{\partial P_{nj}^\nu}
+\kappa \Bigg[\frac{\partial(a^*W)}{\partial a^*}\nonumber\\
&+\frac{\partial(a W)}{\partial a}+(2\bar{n}+1)\frac{\partial^2 W}{\partial a\partial a^*}\Bigg].
\label{eq: Fokker-Planck equation}
\end{align}
Here, we assumed that the electronic excitation energy is sufficiently large compared to the cavity frequency that the electronic ground state is not significantly modified by the light-matter interaction.
The effect of the molecule-cavity interaction on the electronic ground state under the cavity Born-Oppenheimer approximation has been discussed in Refs.~\cite{Fischer23, Schnappinger23, Fiechter24}.
The nuclear dependence of the transition dipole moments to all electronic excited states is also assumed to be small so that the dipole moment fluctuation $\langle \psi_0^\text{e}|\hat{\mu}^2|\psi_0^\text{e}\rangle-\langle \psi_0^\text{e}|\hat{\mu}|\psi_0^\text{e}\rangle^2=\sum_{n\not=0}\langle \psi_0^\text{e}|\hat{\mu}|\psi_n^\text{e}\rangle\langle\psi_n^\text{e}|\hat{\mu}|\psi_0^\text{e}\rangle$ in the electronic ground state, where $|\psi_0^\text{e}\rangle$ and $|\psi_{n\not=0}^\text{e}\rangle$ are the molecule's electronic ground and excited states, respectively, is nearly a constant with respect to the nuclear dynamics and thus can be ignored, and the dipole moment operator $\hat{\mu}$ can be replaced by a dipole moment function $\mu(\hat{\mathbf{r}})=\langle\psi_0^\text{e}|\hat{\mu}|\psi_0^\text{e}\rangle$.

By noticing that a general Fokker-Planck equation $\partial P/\partial t=\left[-\sum_j \partial A_j/\partial x_j+(1/2)\sum_{i,j}\partial^2 D_{ij}/\partial x_i\partial x_j\right]P$ for a distribution function $P(\mathbf{x},t)$ with a positive definite diffusion matrix $D_{ij}(\mathbf{x})$ is equivalent to the Ito stochastic differential equation $\text{d}\mathbf{x}_t=\mathbf{A}(\mathbf{x}_t)\text{d}t+\mathbf{B}(\mathbf{x}_t)\text{d}\mathbf{W}_t$, where $\mathbf{A}(\mathbf{x})$ is the column vector of the drifts $A_j(\mathbf{x})$, the matrix $\mathbf{B}(\mathbf{x})$ is defined by the factorization $\mathbf{D}(\mathbf{x})=\mathbf{B}(\mathbf{x})\mathbf{B}(\mathbf{x})^\text{T}$, and $\mathbf{W}_t$ is a column vector of independent Wiener processes~\cite{Walls-book, Carmichael-book}, from Eq.~\eqref{eq: Fokker-Planck equation}, the stochastic differential equation for the coupled molecule-cavity system is given by 
\begin{align}
\text{d}\mathbf{R}_{nj}=&\frac{\mathbf{P}_{nj}}{M_{nj}}\text{d}t, \label{eq: equation of motion 1} \\ 
\text{d}P_{nj}^\nu=&\Bigg\{-\Bigg(\frac{\partial V_\text{g}^{(n)}}{\partial R_{nj}^\nu}+\sum_{l\not=n}\frac{\partial V_\text{int}^{(nl)}}{\partial R_{nj}^\nu}\Bigg) -\frac{\omega_\text{c}}{\hbar}\Bigg[i\hbar(a^*-a) \nonumber\\
&+2\sum_{n=1}^N \boldsymbol{\mu}_n\cdot\mathbf{A}_0\Bigg]\frac{\partial \mu_n^\lambda}{\partial R_{nj}^\nu}A_0^\lambda\Bigg\}\text{d}t, \\
\text{d}a=&\Bigg[ (-i\omega_\text{c}-\kappa)a+\frac{\omega_\text{c}}{\hbar}\sum_{n=1}^N\boldsymbol{\mu}_n\cdot\mathbf{A}_0\Bigg]\text{d}t+\text{d}A, \label{eq: equation of motion 3}
\end{align}
where $\text{d}A$ represents a complex-number stochastic process with $\text{d}A=\text{d}A_\text{r}+i\text{d}A_\text{i}$ and $\text{d}A_\text{r,i}=\sqrt{\kappa(\bar{n}+1/2)}\text{d}W_\text{r,i}$. 
Here, $\text{d}W_\text{r,i}$ are independent Wiener processes, i.e., $\text{d}W_\text{r,i}=\Theta_\text{r,i}\text{d}t$ and $\Theta_\text{r,i}$ are white noises: $\langle \Theta_{r,i}(t)\rangle=0, \langle \Theta_{r,i}(t)\Theta_{r,i}(t')\rangle=\delta(t-t')$. 
Apart from the decaying and stochastic terms in the last equation, Eqs.~\eqref{eq: equation of motion 1}-\eqref{eq: equation of motion 3} have the same form as the classical equations of motions for the dynamic variables $\mathbf{R}$, $\mathbf{P}$, and $a$~\cite{Li20}.
To the leading order in quantum fluctuation, the expectation value of an arbitrary operator $\hat{\Omega}$ in the coupled molecule-cavity system is given by
\begin{align}
\langle \hat{\Omega}\rangle\simeq &\iint \frac{\text{d}^\mathcal{N}\mathbf{R}_0\text{d}^\mathcal{N}\mathbf{P}_0}{\hbar^\mathcal{N}} \int \text{d}^2 a_0 W_0(\mathbf{R}_0,\mathbf{P}_0,a_0)\nonumber\\
&\times\Omega_\text{W}(\mathbf{R}(t),\mathbf{P}(t),a(t)).
\end{align}
Here, $W_0$ is the Wigner function for the initial density operator $\hat{\rho}(t=0)$, $\Omega_\text{W}$ is the Weyl symbol of the operator $\hat{\Omega}$, and $\mathbf{R}(t)$, $\mathbf{P}(t)$, and $a(t)$ represent the values of the dynamic variables at time $t$ obtained by solving the set of equations of motion~\eqref{eq: equation of motion 1}-\eqref{eq: equation of motion 3} along with the set of initial conditions: $\mathbf{R}(t=0)=\mathbf{R}_0$, $\mathbf{P}(t=0)=\mathbf{P}_0$, and $a(t=0)=a_0$. 

\textit{Molecular-vibration-polariton dynamics in a system of diatomic molecules--}
We now apply the above-developed TWA theory to investigate the quantum dynamics of molecular-vibration-polaritons in a system of identical diatomic molecules. 
The vibration of a molecule in its electronic ground state is modeled by the Morse potential:
\begin{align}
V_\text{g}(q)=D\left(1-e^{-\alpha(q-q_\text{e})}\right)^2.
\end{align} 
Here, $q$ represents the distance between the two nuclei, and $q_\text{e}$ and $D$ denote the equilibrium bond length and the dissociation energy, respectively.
The system is assumed to be sufficiently dilute that the inter-molecular interactions are negligible.
The dipole moment function for a diatomic molecule is assumed to have the following form:
\begin{align}
\mu(q)=Aqe^{-Bq^4}.
\end{align}
In the numerical calculations below, the parameters for the HF molecule will be used: $\alpha=1.174$, $D=0.225$, $q_\text{e}=1.7329$, $A=0.4541$, and $B=0.0064$~\cite{Mondal23}. 
The reduced mass for the relative motion of the two nuclei is $M=1744.59$.
These values are all given in atomic units.

We first consider the initial state that is a tensor-product state in the dipole gauge: $|\psi_0\rangle=|\phi_0\rangle_\text{m}\otimes|0\rangle_\text{c}$, where $|\phi_0\rangle_\text{m}$ is the molecular ground state of the Morse potential and $|0\rangle_\text{c}$ is the vacuum state of the cavity field.
This initial state corresponds to a light-matter entangled state $|\tilde{\psi}_0\rangle=\hat{U}^\dagger |\psi_0\rangle$ in the Coulomb gauge through the inverse PZW transformation.
The Wigner function of the total system at $t=0$ in the dipole gauge is then given by the product of molecular and photonic Wigner functions. 
The molecular Wigner function for the ground state $|\phi_0\rangle_\text{m}$ of the Morse potential is given by~\cite{Dahl88}
\begin{align}
W_\text{m}(q,p)=&\frac{4(2\lambda-1)}{\Gamma(2\lambda)}\left(2\lambda e^{-\alpha(q-q_\text{e})}\right)^{2\lambda-1} \nonumber\\
&\times K_{2ip/(\hbar\alpha)}\left(2\lambda e^{-\alpha(q-q_\text{e})}\right),
\end{align}
where $\lambda=\sqrt{2MD}/\alpha \hbar$, $\Gamma(z)=\int_0^\infty\text{d}t\, t^{z-1}e^{-t}$ is the gamma function, and $K_\nu(\xi)=(1/2)\int_0^\infty\text{d}x\, x^{\nu-1}\exp\left[-(\xi/2)(x+1/x)\right]$ is the modified Bessel function of the third kind.
By making a variable transformation $u=\ln x$, the modified Bessel function can be rewritten as $K_\nu(\xi)=(1/2)\int_{-\infty}^\infty\text{d}u\,\exp\left(\nu u-\xi\cosh u\right)$ and calculated numerically using, for example, the Simpson's 1/3 formula~\cite{Gilat-book}.
The photonic Wigner function for the vacuum state of cavity photons is given by $W_\text{c}(a)=(2/\pi)e^{-2|a|^2}$.
An ensemble of initial values of $q$, $p$, and $a$ following the Wigner distribution function is created by using the Monte Carlo M(RT)$^2$ algorithm~\cite{Tuckerman-book}. 
Note that through the PZW transformation, the molecular operator $\hat{q}$ is unchanged while the cavity photon number $\hat{n}=\hat{a}^\dagger\hat{a}$ is transformed to 
\begin{align}
\hat{\tilde{n}}=&\hat{U}\hat{n}\hat{U}^\dagger \nonumber\\
=&\hat{a}^\dagger\hat{a}+\frac{i}{\hbar}A_0\sum\limits_{j=1}^N\mu(\hat{q}_j)\left(\hat{a}^\dagger-\hat{a}\right)+\left(\frac{A_0}{\hbar}\sum\limits_{j=1}^N\mu(\hat{q}_j)\right)^2,
\end{align} 
where $A_0$ is the projection of $\mathbf{A}_0$ on the molecular direction.
The corresponding Weyl symbol of $\hat{\tilde{n}}$ is given by 
\begin{align}
\tilde{n}_\text{W}=|a|^2-\frac{1}{2}+\frac{2A_0}{\hbar}\text{Im}\left\{a\right\}\sum\limits_{j=1}^N\mu(q_j)+\left(\frac{A_0}{\hbar}\sum_{j=1}^N\mu(q_j)\right)^2.
\end{align}

In the numerical calculations below, the molecule-cavity collective coupling strength is set to $g\sqrt{N}A\sqrt{\frac{\hbar}{M\omega_0}}=0.1\hbar\omega_\text{c}$, corresponding to the ultrastrong coupling regime~\cite{Kockum19}.
Here, $g=A_0\omega_\text{c}$ characterizes the light-matter coupling strength, $\omega_0=\alpha\sqrt{2D/M}$ is the harmonic-oscillator frequency at the equilibrium point of the Morse potential, and $\sqrt{\frac{\hbar}{M\omega_0}}$ is the characteristic length scale of a harmonic oscillator with frequency $\omega_0$. 
As a result of the ultrastrong coupling, both the molecular nuclei and cavity field would evolve with time, even if they are initially prepared in their ground states.
The cavity frequency $\omega_\text{c}$ is chosen to be resonant with the energy difference $E_{01}=E_1-E_0$ between the vibrational ground state $|\phi_0\rangle_\text{m}$ and the first excited state $|\phi_1\rangle_\text{m}$ of the Morse potential. 
The energy eigenvalues for the Morse potential is given by $E_n=\left[(n+1/2)-(n+1/2)^2/(2\lambda)\right]\hbar\omega_0$ for $n=0,1,\cdots, [\lambda-1/2]$, where $[x]$ denotes the largest integer smaller than $x$~\cite{Dahl88}.
The cavity loss rate $\kappa$ is chosen to correspond to an energy of $10\;\text{meV}$.

\begin{figure}[htbp] % float placement: (h)ere, page (t)op, page (b)ottom, other (p)age
  \centering
  % file name: E:/Draft-Semiclassical TWA theory of molecular vibration polariton dynamics (Sep 2023)/compare_semiclassical_vs_quantum_v2.eps
  \includegraphics[width=3.4in,keepaspectratio]{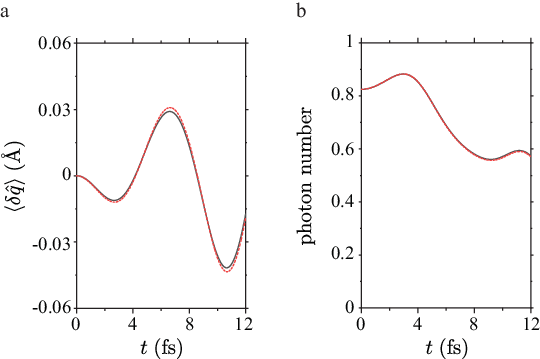}
  \caption{Time evolutions of the expectation values of (a) the molecular nuclear displacement $\delta\hat{q}=\hat{q}-q_0$ and (b) the cavity photon number obtained by the semiclassical TWA (red, dashed) and the fully quantum mechanical calculation (black, solid) for the initial state that is a tensor product of the molecular vibrational ground state and the photonic vacuum state in the dipole gauge and corresponds to a light-matter entangled state in the Coulomb gauge. Here, $q_0$ is the initial expectation value of the nuclear distance operator $\hat{q}$. The coupling strength is in the ultrastrong coupling regime such that both the molecular nuclei and the cavity field evolve with time, even if they are initially prepared in their ground states.}
  \label{fig: compare semiclassical vs quantum}
\end{figure}

We first consider the single-molecule system, i.e., $N=1$, for which a fully quantum mechanical calculation is doable. 
Figure~\ref{fig: compare semiclassical vs quantum} compares the molecular nuclear dynamics and time evolution of the cavity photon number obtained by the TWA with those obtained by the fully quantum mechanical theory.
The fully quantum mechanical result is obtained by solving the quantum master equation~\eqref{eq: quantum master equation}.
The wavefunction of the ground state of the Morse potential is given by~\cite{Dahl88}
\begin{align}
\phi_0(q)=&\sqrt{\frac{(2\lambda-1)\alpha}{\Gamma(2\lambda)}}\left[2\lambda e^{-\alpha(q-q_\text{e})}\right]^{\lambda-1/2} \nonumber\\
&\times \exp\left[-\lambda e^{-\alpha(q-q_\text{e})}\right].
\end{align}
A good agreement between the semiclassical TWA and fully quantum mechanical results is observed for the dynamics of both molecular nuclei and cavity photons.
Note that the expectation value of the cavity photon number is nonzero at $t=0$ since the initial state corresponds to a light-matter entangled state in the Coulomb gauge.

\begin{figure}[tbp] % float placement: (h)ere, page (t)op, page (b)ottom, other (p)age
  \centering
  % file name: E:/Draft-Semiclassical TWA theory of molecular vibration polariton dynamics (Sep 2023)/compare_collective.eps
  \includegraphics[width=2.8in,keepaspectratio]{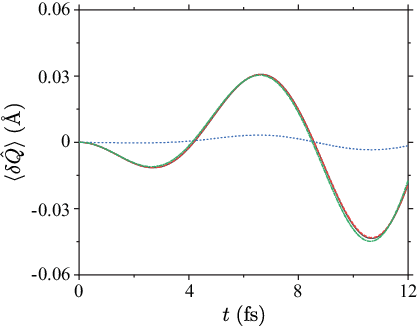}
  \caption{Time evolution of the expectation value of the nuclear displacement operator averaged over all the molecules inside the cavity $\delta\hat{Q}=\hat{Q}-q_0=(1/N)\sum_{j=1}^N \hat{q}_j-q_0$ for a system of $N=30$ identical diatomic molecules, all aligned parallel to the polarization direction of the cavity mode (black, solid). 
For comparison, the nuclear dynamics of a single-molecule system ($N=1$) with constant collective coupling strength (red, dash), that with constant coupling strength per molecule (blue, dot), and the fully quantum dynamics for a system of $N=2$ molecules with constant collective coupling strength (green, dash dot) are also shown. 
The initial state is a tensor-product ground state in the dipole gauge as in Fig.~\ref{fig: compare semiclassical vs quantum}.}
  \label{fig: compare collective}
\end{figure}

We next consider a system of $N=30$ diatomic molecules, all aligned parallel to the polarization direction of the cavity mode. 
Here, the molecule-cavity collective coupling strength is kept constant, i.e., $g\sqrt{N}A\sqrt{\frac{\hbar}{M\omega_0}}=0.1\hbar\omega_\text{c}$, implying that the coupling strength per molecule is reduced by a factor of $1/\sqrt{N}$ from the above case of a single-molecule system. 
The initial state is a tensor-product ground state in the dipole gauge: $|\psi_0\rangle=\prod_{j=1}^N|\phi_0^{(j)}\rangle_\text{m}\otimes |0\rangle_\text{c}$.
Figure~\ref{fig: compare collective} shows the nuclear dynamics averaged over all the molecules inside the cavity.
For comparison, the nuclear dynamics of the single-molecule system ($N=1$) with constant collective coupling strength, i.e., $gA\sqrt{\frac{\hbar}{M\omega_0}}=0.1\hbar\omega_\text{c}$, and that with constant coupling strength per molecule, i.e., $gA\sqrt{\frac{\hbar}{M\omega_0}}=0.1\hbar\omega_\text{c}/\sqrt{N}$ are also shown in the figure.
It is evident that the amplitude of the averaged nuclear dynamics does not decrease with the increasing number of molecules as long as the collective coupling strength is kept constant, and it is much larger than that of the single-molecule system with the same coupling strength per molecule. 
The validity of the TWA is also confirmed to be not limited to a single-molecule system by looking at the fully quantum dynamical result for a system of $N=2$ molecules with constant collective coupling strength shown in Fig.~\ref{fig: compare collective}. 
It is clear that the nuclear dynamics is not reduced by the increasing number of molecules inside the cavity, as opposed to the circumstance of a group of harmonic oscillators linearly coupled to an optical cavity, where as long as the collective light-matter coupling strength is kept constant, the dynamic amplitude of the collective mode $\hat{Q}_\text{col}=(1/\sqrt{N})\sum_{j=1}^N \hat{q}_j$ is independent of the number of oscillators, and thus, the dynamic amplitude of an averaged observable such as $\hat{Q}=(1/N)\sum_{j=1}^N \hat{q}_j$ is strongly suppressed in large systems.

\begin{figure}[tbp] % float placement: (h)ere, page (t)op, page (b)ottom, other (p)age
  \centering
  % file name: E:/Draft-Semiclassical TWA theory of molecular vibration polariton dynamics (Sep 2023)/resonance_v2.eps
  \includegraphics[width=2.8in,keepaspectratio]{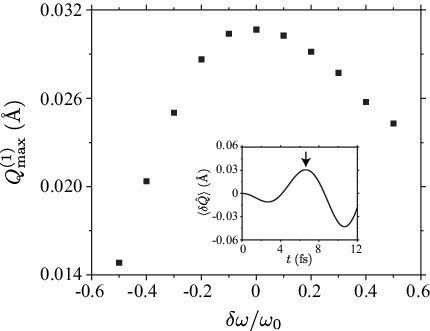}
  \caption{The first local maximum $Q_\text{max}^{(1)}$ (indicated by the arrow in the inset) in the time evolution of the expectation value $\langle\delta\hat{Q}\rangle$ of the averaged nuclear displacement operator as a function of the detuning frequency $\delta\omega=\omega_\text{c}-E_{01}/\hbar$ normalized by $\omega_0$.}
  \label{fig: resonance}
\end{figure}

\begin{figure}[tbp] % float placement: (h)ere, page (t)op, page (b)ottom, other (p)age
  \centering
  % file name: E:/Draft-Semiclassical TWA theory of molecular vibration polariton dynamics (Sep 2023)/compare_cavity_loss.eps
  \includegraphics[width=2.8in,keepaspectratio]{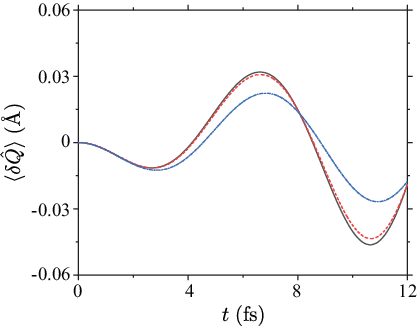}
  \caption{Time evolutions of the expectation value $\langle\delta\hat{Q}\rangle$ of the averaged nuclear displacement operator for different values of the cavity loss rates $\kappa$ corresponding to $1\;\text{meV}$ (black, solid), $10\;\text{meV}$ (red, dashed), and $100\;\text{meV}$ (blue, dashed dot).}
  \label{fig: compare cavity loss}
\end{figure}

To investigate the dependence of the molecular nuclear dynamics on the cavity frequency, the first local maximum $Q_\text{max}^{(1)}$ in the time evolution of the expectation value $\langle\delta\hat{Q}\rangle$ of the averaged nuclear displacement operator is shown in Fig.~\ref{fig: resonance} as a function of the detuning frequency $\delta\omega=\omega_\text{c}-E_{01}/\hbar$.
Here, $\delta\hat{Q}=\hat{Q}-q_0$ with $q_0=\langle\phi_0|\hat{q}|\phi_0\rangle$.
It is clear that $Q_\text{max}^{(1)}$ is maximum when the cavity frequency is resonant with the energy difference between the vibrational ground state and the first excited state of the Morse potential. 
The dependence of the molecular nuclear dynamics on the cavity loss rate $\kappa$ is also investigated. 
Figure~\ref{fig: compare cavity loss} shows the nuclear dynamics for the rates corresponding to $1$, $10$, and $100\;\text{meV}$. 
The dynamic amplitude of the averaged nuclear displacement decreases with the increasing cavity loss rate as a result of the energy dissipation. 

\begin{figure}[tbp] % float placement: (h)ere, page (t)op, page (b)ottom, other (p)age
  \centering
  % file name: E:/Draft-Semiclassical TWA theory of molecular vibration polariton dynamics (Sep 2023)/random_orientation.eps
  \includegraphics[width=3.4in, keepaspectratio]{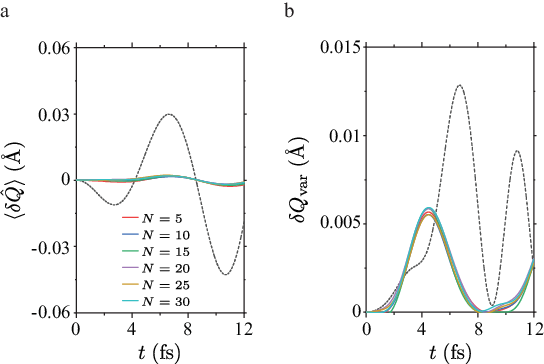}
  \caption{Time evolutions of (a) the expectation value $\langle \delta\hat{Q}\rangle$ of the averaged nuclear displacement operator and (b) the square root of the averaged variation of the nuclear displacement $\delta Q_\text{var}$ whose definition is given in the main text. The molecular orientations are distributed evenly in space. The number of molecules varies $N=5, 10, 15, 20, 25, 30$ (indicated by the colors) while the collective coupling strength is kept constant. The nuclear dynamics of a single-molecule system ($N=1$) with constant collective coupling strength (black, dashed) are shown in both (a) and (b) for comparison.}
  \label{fig: random orientation}
\end{figure}

We next investigate the nuclear dynamics in a system of molecules with random orientations strongly coupled to an optical cavity mode. 
The numerical calculations below consider a system of $N$ diatomic molecules whose orientations are distributed evenly in the range of angles $0\leq \theta\leq \pi$, i.e., $\theta_j=(j-1)\pi/(N-1) (j=1,\cdots,N)$. 
Here, $\theta_j$ is the angle between the direction of the $j$th molecule and the polarization direction of the cavity mode. 
As the molecule-cavity coupling $g_j$ for the $j$th molecule is proportional to $\cos\theta_j$, its sign changes as the angle crosses $\pi/2$.
Figure~\ref{fig: random orientation}a shows the time evolutions of the expectation value of the nuclear displacement operator averaged over all the molecules $\delta\hat{Q}$ for systems with varying numbers of molecules. 
It can be seen that $\langle \delta\hat{Q}\rangle$ for a system of many molecules with random orientations is small compared to that for the single-molecule system with constant collective coupling strength. 
This is due to the dependence of the sign of the molecule-cavity coupling on the molecule's orientation. 
For example, in a system of two diatomic molecules with opposite directions, we have $g_2=-g_1$. 
The cavity mode then only effectively couples to the nuclear motion's antisymmetric mode $\hat{Q}_\text{as}=(\hat{q}_1-\hat{q}_2)/\sqrt{2}$ rather than the symmetric mode $\hat{Q}_\text{s}=(\hat{q}_1+\hat{q}_2)/\sqrt{2}$. 
As a result, as long as such an observable that is a linear function of $\hat{Q}_\text{s}$ as the averaged nuclear displacement is considered, a small dynamic amplitude should be expected. 
However, a larger dynamic amplitude can be observed for an observable that is a nonlinear function of $\hat{Q}_\text{s}$.
Figure~\ref{fig: random orientation}b shows the time evolution of the square root of the averaged variation of the nuclear displacement defined by $\delta Q_\text{var}(t)=\sqrt{\langle\hat{Q}_\text{var}\rangle(t)}-\sqrt{\langle\hat{Q}_\text{var}\rangle(t=0)}$, where $\hat{Q}_\text{var}=(1/N)\sum_{j=1}^N (\hat{q}_j-q_0)^2$.
It is evident that the dynamic amplitude of $\delta Q_\text{var}$ for a system of many molecules with different orientations is smaller than that for the single-molecule system with constant collective coupling strength by a factor that is comparable to the reduction of the averaged coupling strength due to random orientations $\overline{\cos^2\theta_j}\simeq1/2$.
Moreover, the dynamic amplitude of $\delta Q_\text{var}$ does not decrease with the increasing number of molecules. 
The above collective effect of molecular polariton formation on the nuclear dynamics of a system containing many molecules can be attributed to the nonzero molecular dipole moment $\mu(q_e)\not=0$ at the equilibrium point as the light-matter interaction in the dipole-gauge Hamiltonian (Eq.~\eqref{eq: molecule-cavity coupling}) involves the sum of dipole moments over all the molecules. 
Equivalently, the collective dynamic behavior can be understood as a consequence of the initial light-matter entangled state 
\begin{align}
|\tilde{\psi}_0\rangle=&\hat{U}^\dagger|\psi_0\rangle \nonumber\\
=&\exp\left\{i\frac{A_0}{\hbar}\left(\hat{a}^\dagger+\hat{a}\right)\sum\limits_{j=1}^N \mu(\hat{q}_j)\right\}\left[\prod\limits_{j=1}^N|\phi_0^{(j)}\rangle_\text{m}\otimes|0\rangle_\text{c}\right],
\end{align}
which also involves the sum of molecular dipole moments.

\begin{figure}[tbp] % float placement: (h)ere, page (t)op, page (b)ottom, other (p)age
  \centering
  % file name: E:/Draft-Semiclassical TWA theory of molecular vibration polariton dynamics (Sep 2023)/compare_transformed_semiclassical_vs_quantum.eps
  \includegraphics[width=3.4in, keepaspectratio]{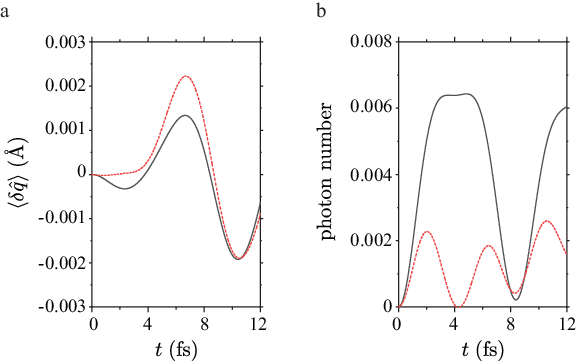}
  \caption{Time evolutions of the expectation values of (a) the nuclear displacement $\delta\hat{q}$ and (b) the cavity photon number obtained by the semiclassical TWA (red, dashed) and the fully quantum mechanical calculation (black, solid) in a single-molecule system ($N=1$) for the initial state that is a tensor product of the molecular vibrational ground state and the photonic vacuum state in the Coulomb gauge and corresponds to a light-matter entangled state in the dipole gauge.}
  \label{fig: compare semiclassical vs quantum in the transformed Coulomb gauge}
\end{figure}

Finally, we investigate the quantum dynamics of the strongly coupled molecule-cavity system for the initial state that is a tensor product of the molecular vibrational ground state and the photonic vacuum state in the Coulomb gauge and corresponds to a light-matter entangled state in the dipole gauge. 
The Wigner function of the initial state in the dipole gauge is no longer separable and it is calculated to be
\begin{widetext}
\begin{align}
W(\{q_j\},\{p_j\},a)=&\int\limits_{-\infty}^\infty\cdots \int\limits_{-\infty}^\infty \prod\limits_{j=1}^N\text{d}\xi_j \;e^{\frac{i}{\hbar}\sum\limits_{j=1}^N p_j\xi_j} 
\prod\limits_{j=1}^N\left[\phi_0\left(q_j-\frac{\xi_j}{2}\right)\phi_0^*\left(q_j+\frac{\xi_j}{2}\right)\right]
\int\text{d}^2\eta\; e^{\eta^*a-\eta a^*}
\langle 0|\hat{U}_+^\dagger e^{\eta \hat{a}^\dagger-\eta^*\hat{a}} \hat{U}_-|0\rangle
\nonumber\\
=&\;2e^{-2|a|^2}\left[\frac{(2\lambda-1)\alpha}{\Gamma(2\lambda)}\right]^N \left[(2\lambda)^N e^{-\alpha\sum\limits_{j=1}^N (q_j-q_\text{e})}\right]^{2\lambda-1}
\int\limits_{-\infty}^\infty\cdots \int\limits_{-\infty}^\infty \prod\limits_{j=1}^N\text{d}\xi_j \;e^{\frac{i}{\hbar}\sum\limits_{j=1}^N p_j\xi_j}\nonumber\\
&\times\exp\left\{-2i\frac{A_0}{\hbar}\sum\limits_{j=1}^N \left[a^*\mu\left(q_j+\frac{\xi_j}{2}\right)-a\mu\left(q_j-\frac{\xi_j}{2}\right)\right]
-2\lambda \sum\limits_{j=1}^N \left[e^{-\alpha(q_j-q_\text{e})}\cosh\left(\frac{\alpha\xi_j}{2}\right)\right]\right.\nonumber\\
&\left.-\frac{1}{2}\left(\frac{A_0}{\hbar}\right)^2 \left[\sum\limits_{j=1}^N \mu\left(q_j+\frac{\xi_j}{2}\right)+\sum\limits_{j=1}^N\mu\left(q_j-\frac{\xi_j}{2}\right)\right]^2\right\},
\label{eq: transformed Wigner function}
\end{align}
\end{widetext}
where $\hat{U}_\pm=\exp\left\{-\frac{i A_0}{\hbar}\sum\limits_{j=1}^N\mu\left(q_j\pm\frac{\xi_j}{2}\right)(\hat{a}+\hat{a}^\dagger)\right\}$.
Here, in the derivation of the second equality in Eq.~\eqref{eq: transformed Wigner function}, we used the following properties of the optical coherent states and displacement operators: $\hat{D}(\alpha)=e^{\alpha\hat{a}^\dagger-\alpha^*\hat{a}}$, $\hat{D}(\alpha)|\beta\rangle=|\alpha+\beta\rangle e^{i\text{Im}\{\alpha\beta^*\}}$, and $\langle \alpha|\beta\rangle=e^{-|\alpha-\beta|^2/2}e^{i\text{Im}\{\alpha^*\beta\}}$.
Due to the complexity of the Wigner function of a light-matter entangled state, we make a transformation back to the Coulomb gauge where the initial state is a tensor product state.
The Hamiltonian is then transformed to
\begin{align}
\hat{\tilde{H}}=&\hat{U}^\dagger \hat{H}\hat{U}\nonumber\\
=&\hbar\omega_\text{c}\left(\hat{a}^\dagger\hat{a}+\frac{1}{2}\right)\nonumber\\
&+\sum\limits_{j=1}^N \left\{\frac{\left[\hat{p}_j-A_0\left(\hat{a}^\dagger+\hat{a}\right)\mu'(\hat{q}_j)\right]^2}{2M}+V(\hat{q}_j)\right\}.
\end{align}
Here, we used the identity $e^{\hat{X}}\hat{Y}e^{-\hat{X}}=\hat{Y}+[\hat{X},\hat{Y}]+[\hat{X},[\hat{X},\hat{Y}]]/2!+\cdots$. 
As a nonzero molecular dipole moment $\mu(q_e)$ at the equilibrium point becomes irrelevant in the Hamiltonian $\hat{\tilde{H}}$ through the first derivative $\mu'(\hat{q})$, it is likely that the collective dynamic behavior would not be observed for the initial tensor product state in the Coulomb gauge.
The Weyl symbol of the Hamiltonian $\hat{\tilde{H}}$ is given by
\begin{align}
\tilde{H}_\text{W}=\hbar\omega_\text{c}|a|^2 +\sum\limits_{j=1}^N \left\{\frac{\left[p_j-2A_0\text{Re}\{a\} \mu'(q_j)\right]^2}{2M}+V(q_j)\right\}
\end{align}
Here, we used the fact that the Weyl symbols of the operators $\hat{p}_j\mu'(\hat{q}_j)$ and $\mu'(\hat{q}_j)\hat{p}_j$ are $p_j\mu'(q_j)\mp (i\hbar/2)\mu''(q_j)$, respectively.
Figure~\ref{fig: compare semiclassical vs quantum in the transformed Coulomb gauge} compares the molecular nuclear dynamics and the time evolution of the cavity photon number obtained by using the semiclassical TWA with $\tilde{H}_\text{W}$ with those obtained by a fully quantum mechanical calculation.
Here, for simplicity we set $N=1$ and $\kappa=0$, while the other parameters are the same as in the previous numerical calculations.
A noticeable difference between the semiclassical TWA and quantum mechanical results is observed in the dynamics of the strongly coupled molecule-cavity system for the initial tensor-product ground state in the Coulomb gauge.
This can be understood as a consequence of the strong light-matter entanglement.
The dependence of the validity of TWA on whether the initial state is a tensor-product state in the dipole or Coulomb gauge implies that the dynamics of the molecular system and cavity field are more entangled with each other in the Coulomb gauge than in the dipole gauge.
This is consistent with the fact that in finding the energy eigenvalues of the hybrid system, the Hamiltonian in the Coulomb gauge often requires a larger set of basis states to converge than the Hamiltonian in the dipole gauge~\cite{Bernardis18}.

\textit{Conclusion--}
We have developed the semiclassical TWA theory of molecular-vibration-polariton dynamics that is tractable in large molecular systems while simultaneously capturing the quantum character of photons in the optical cavity. 
The theory was then applied to investigate the nuclear quantum dynamics of a system of identical diatomic molecules with a ground-state Morse potential strongly coupled to an infrared cavity mode in the ultrastrong coupling regime. 
For the initial state that is a tensor product of the molecular vibrational ground state and the photonic vacuum state in the dipole gauge and corresponds to a light-matter entangled state in the Coulomb gauge, a good agreement between the semiclassical TWA and quantum mechanical results was observed for the dynamics of the coupled molecule-cavity system. 
The collective and resonance effects were also observed in the nuclear dynamics of a system containing many molecules.
By contrast, for the initial tensor-product ground state in the Coulomb gauge corresponding to a light-matter entangled state in the dipole gauge, a noticeable difference between the semiclassical and quantum mechanical dynamics is observed as a consequence of the strong light-matter entanglement.
For a general initial state, the TWA is expected to be more valid for thermally activated dynamical processes and/or molecular systems with close-to-classical behaviors as the light-matter entanglement can be suppressed by dephasing.  
A possible future direction is to develop the TWA theory for the molecular-exciton-polariton dynamics, by which various interesting phenomena including, for example, the effect of polaron decoupling~\cite{Spano15, Herrera16, Phuc19, Takahashi20}, super-reaction~\cite{Phuc21}, Bose enhancement of excitation-energy transport~\cite{Phuc22}, and photon-coupled electron spin dynamics~\cite{Phuc23}, can be quantitatively studied in large molecular systems.

%%%%%%%%%%%%%%%%%%%%%%%%%%
%\section{Supplementary Material}
%\label{sec: Supplementary Material}
%See Supplementary Material for.

%%%%%%%%%%%%%%%%%%%%%%%%%%
\begin{acknowledgements}
N. T. Phuc would like to thank Pham Quang Trung for fruitful discussion on numerical calculations. 
We acknowledge financial support from Hirose Foundation.
The computations were performed using Research Center for Computational Science, Okazaki, Japan.
\end{acknowledgements}

%%%%%%%%%%%%%%%%%%%%%%%%%%
%\section*{Data availability}
%The data that support the findings of this study are available from the corresponding author upon reasonable request.

%%%%%%%%%%%%%%%%%%%%%%%%%%

%%%%%%%%%%%%%%%
\end{document}